\documentclass{PoS}

\usepackage{amsmath}
\usepackage{amssymb}

\title{Hadroproduction of open heavy flavour for PDF analyses}

\ShortTitle{Hadroproduction of open heavy flavour for PDF analyses}

\author{Ilkka Helenius \\
        University of Jyvaskyla, Department of Physics, P.O. Box 35, FI-40014 University of Jyvaskyla, Finland \\
        Helsinki Institute of Physics, P.O. Box 64, FI-00014 University of Helsinki, Finland \\
        Institute for Theoretical Physics, T\"ubingen University, Auf der Morgenstelle 14, 72076 T\"ubingen, Germany \\
        E-mail: \email{ilkka.m.helenius@jyu.fi}}

\author{\speaker{Hannu Paukkunen}\\
        University of Jyvaskyla, Department of Physics, P.O. Box 35, FI-40014 University of Jyvaskyla, Finland \\
        Helsinki Institute of Physics, P.O. Box 64, FI-00014 University of Helsinki, Finland \\
        E-mail: \email{hannu.paukkunen@jyu.fi}}

\abstract{
Due to the large masses of the charm and bottom quarks, their production cross sections are calculable within the perturbative QCD. This makes the heavy-quark mesons important observables in high-energy collisions of protons and nuclei. However, the available calculations for heavy-flavored-meson hadroproduction have been somewhat problematic in reliably describing the cross sections across the full kinematic range from zero to very high $p_{\rm T}$. This has put some question marks on the robustness of LHC heavy-flavored-meson measurements in studying the partonic structure of the colliding hadrons and nuclei. Here, we introduce SACOT-$m_{\rm T}$ -- a novel scheme for open heavy-flavour hadroproduction within the general-mass variable-flavour-number formalism that solves this problem. The introduced scheme is an analogue of the SACOT-$\chi$ scheme often used for deeply-inelastic scattering in global analyses of PDFs.
}

\FullConference{XXVII International Workshop on Deep-Inelastic Scattering and Related Subjects - DIS2019\\
		8-12 April, 2019\\
		Torino, Italy}

\begin{document}

\section{Introduction}
\vspace{-0.2cm}

The hadroproduction of open heavy flavour -- D mesons in particular -- has recently been advocated as a promising constraint for proton parton distribution functions (PDFs) \cite{Zenaiev:2015rfa,Gauld:2016kpd}. The theoretical description is typically based on the fixed flavour-number scheme (FFNS) \cite{Mangano:1991jk}, \textsc{fonll} code \cite{Cacciari:1998it}, or FFNS matched to parton showers \cite{Frixione:2007nw}. However, the use of e.g. FFNS calculation in conjunction with PDFs defined in variable flavour-number schemes (the commonly used general-purpose PDFs) may be too restrictive, and in this sense calculations within the framework of general-mass variable-flavour-number scheme (GM-VFNS) would be more natural. Here, we will discuss our novel implementation of the GM-VFNS, the so-called SACOT-$m_{\rm T}$ scheme \cite{Helenius:2018uul}. 

\vspace{-0.2cm}
\section{D-meson production in fixed flavour-number scheme}
\vspace{-0.2cm}

Within FFNS -- assuming no intrinsic heavy-quark content in the proton -- the massive quarks $Q$ are always produced in pairs by the partonic processes, $g+g \rightarrow Q\overline{Q} + X, \, q+\overline{q} \rightarrow Q\overline{Q} + X, \, q+g \rightarrow Q\overline{Q} + X\,$. The cross section for inclusive $Q$ production can be written as an integral of PDFs $f_i^{p}(x_1,\mu^2_{\rm fact})$ and partonic cross sections $d\hat\sigma^{ij}$,
\begin{align}
\frac{d\sigma(p + p \rightarrow Q + X)}{dp_{\rm T}dy} = & \sum _{ij}
\int dx_1 dx_2 
         { f_i^{p}(x_1,\mu^2_{\rm fact}) }
            { \frac{d\hat{\sigma}^{ij\rightarrow Q + X}(\tau_1, \tau_2, m^2, \mu^2_{\rm ren}, \mu^2_{\rm fact})}{dp_{\rm T} dy} }
         { f_j^{p}(x_2,\mu^2_{\rm fact}) }  \,,           
        \nonumber
\end{align}
where $y$ and $p_{\rm T}$ denote the rapidity and transverse momentum of the produced heavy quark. The factorization and renormalization scales are marked by $\mu^2_{\rm fact}$ and $\mu^2_{\rm ren}$. The kinematic variables $\tau_{1,2}$ are the ``massive'' Mandelstam variables,
$$\tau_1 \equiv {p_1 \cdot p_3}/{p_1 \cdot p_2} = {m_{\rm T}e^{-y}}/({\sqrt{s} x_2}), \ \ \ \
\tau_2 \equiv {p_2 \cdot p_3}/{p_1 \cdot p_2} = {m_{\rm T}e^{y}}/({\sqrt{s} x_1}), \ \ \ \ m^2_{\rm T} = {p_{\rm T}^2+m^2} \,, $$
denoting the momenta of the incoming partons and outgoing heavy quark by $p_{1,2}$ and $p_3$, respectively. The partonic cross sections scale as $d\hat{\sigma}^{ij\rightarrow Q + X} \sim  \tau_{1,2}^{-n}$, and, thanks to the heavy-quark mass, remain finite even at $p_{\rm T}=0$. The heavy-quark cross sections can be turned into, say D$^0$-meson production cross sections by folding with $Q\rightarrow \mathrm{D^0}$ fragmentation functions (FFs) ${D_{Q \rightarrow \mathrm{D}^0}(z)}$. The fragmentation variable $z$ is not unique when the masses of the heavy quark and D$^0$ meson are kept non zero. For simplicity, we define $z \equiv {E_{\rm D^0}}/{E_Q}$, where $E_{\rm D^0}$ and $E_Q$ are the energies of the D$^0$ meson and heavy quark in the center-of-mass frame of the p-p collision. Assuming that the fragmentation is collinear we get,
\begin{align}
\frac{d\sigma(p + p \rightarrow \mathrm{D}^0 + X)}{dP_{\rm T}dY} = & \sum _{ij}
\int \frac{dz}{z} dx_1 dx_2
            { f_i^{p}(x_1,\mu^2_{\rm fact}) }
            {\frac{d\hat{\sigma}^{ij\rightarrow Q + X}}{dp_{\rm T} dy} }
            {f_j^{p}(x_2,\mu^2_{\rm fact})}             
            {D_{Q \rightarrow \mathrm{D}^0}(z) }             
            \nonumber \,,
\end{align}
where the (lower case) partonic and (upper case) hadronic variables are related by
\begin{align}
p_{\rm T}^2 & = \frac{M^2_{\rm T}\cosh^2 Y - z^2m^2}{z^2} \left(1 + \frac{M_{\rm T}^2\sinh^2 Y}{P^2_{\rm T}}\right)^{-1} \xrightarrow{P_{\rm T} \rightarrow \infty} \left(\frac{P_{\rm T}}{z}\right)^2 \nonumber \,,  \\ 
    y & = \sinh^{-1} \left(\frac{M_{\rm T} \sinh Y}{P_{\rm T}} \frac{p_{\rm T}}{m_{\rm T}}\right) \xrightarrow{P_{\rm T} \rightarrow \infty} Y \,,  \nonumber
\end{align}
and $M_{\rm T} = \sqrt{P_{\rm T}^2+M_Q^2}$ marks the hadronic transverse mass. While this framework appears to work well at low $P_{\rm T}$ (see e.g. \cite{Zenaiev:2016kfl}), the description seems to deteriorate towards high $P_{\rm T}$. Presumably, this can be attributed to the $\log(p_{\rm T}^2/m^2)$ behaviour of the partonic cross sections which begin to dominate and should be resummed, as we will come to conlcude.

\vspace{-0.2cm}
\section{D-meson production in general-mass variable-flavour-number scheme}
\vspace{-0.2cm}

\begin{figure}[htb!]
\centering
\includegraphics[width=0.90\linewidth]{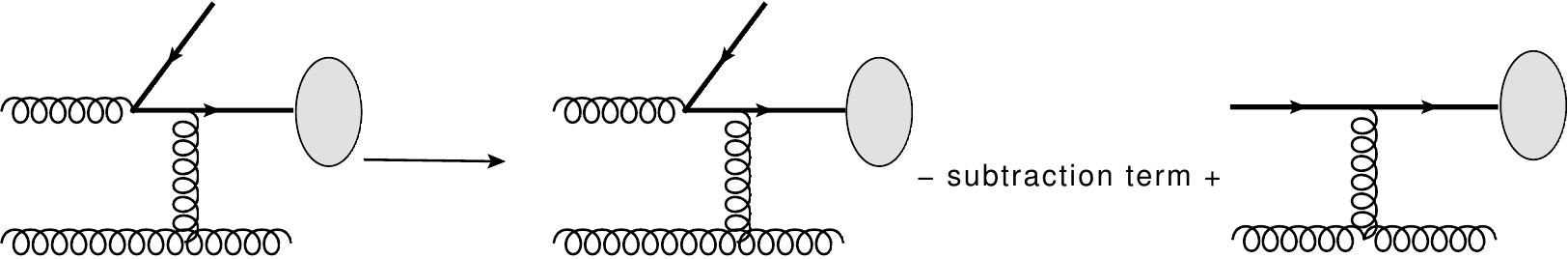}
\\
\hspace{-0.3cm} (a) \hspace{4.5cm} (a) \hspace{2.5cm} (b) \hspace{2.0cm} (c)
\caption{Origin of the heavy-quark initiated subprocess illustrated.}
\label{fig:ini}
\end{figure}

The GM-VFNS description can be obtained from FFNS by resumming the $\log(p_{\rm T}^2/m^2)$ terms that appear in FFNS. As an example, in Figure~\ref{fig:ini} the diagram (a) gives rise to such logarithmic behaviour when the initial-state gluon splits into a $Q\overline{Q}$ pair. This is only the first of the whole series of diagrams that are in GM-VFNS summed into the heavy-quark PDF $f_Q^{p}$.
Effectively, this summation can be realized by including the heavy-quark initiated contribution (c) and a subtraction term (b) that avoids the double counting between diagrams (a) and (c). The contribution from $Qg\rightarrow Q+X$ channel, represented here by the diagram (c), can be written as
\begin{align}
\int \frac{dz}{z}dx_1 dx_2 \,
            { f_Q^{p}(x_1,\mu^2_{\rm fact}) }
            {\frac{d\hat{\sigma}^{Qg\rightarrow Q + X}(\tau_1, \tau_2)}{dp_{\rm T} dy} }
            { f_g^{p}(x_2,\mu^2_{\rm fact}) }             
            {D_{Q \rightarrow \mathrm{D}^0}(z)}
        \nonumber \,.
\end{align}
The subtraction term (b) is obtained from this same expression by replacing the heavy-quark PDF with its perturbative expression,
\vspace{-0.2cm}
\begin{equation}
f^p_Q(x,\mu^2_{\rm fact}) = \left(\frac{\alpha_s}{2\pi}\right) \log\left(\frac{\mu^2_{\rm fact}}{m^2}\right) \int_x^1 \frac{\mathrm{d}\ell}{\ell} P_{qg}\left(\frac{x}{\ell}\right) f_g^p(\ell,\mu^2_{\rm fact}) + \mathcal{O}(\alpha_s^2) \,, \nonumber
\end{equation}
where $\alpha_s$ is the QCD coupling, and $P_{qg}$ is the usual gluon-to-quark splitting function. However, the exact form of $d\hat{\sigma}^{Qg\rightarrow Q + X}(\tau_1, \tau_2)$ in the above expressions is not unique \cite{Thorne:2008xf}. In practice, we can only require that the zero-mass $\overline{\rm MS}$ expressions are recovered at high $p_{\rm T}$,
\begin{align}
            {\frac{d\hat{\sigma}^{Qg\rightarrow Q + X}(\tau_1, \tau_2)}{dp_{\rm T} dy}}
        \xrightarrow{p_{\rm T} \rightarrow \infty}
            {\frac{d\hat{\sigma}^{qg\rightarrow q + X}(\tau_1^0, \tau_2^0)}{dp_{\rm T} dy}} \,,
        \ \ \ \ \ \ \tau_{1,2}^0 = \tau_{1,2} \xrightarrow{m \rightarrow 0} {p_{\rm T}e^{\mp y}}/({\sqrt{s} x_{2,1}}) \,.
\nonumber
\end{align}
The easiest option is to define ${d\hat{\sigma}^{Qg\rightarrow Q + X}(\tau_1, \tau_2)} \equiv {d\hat{\sigma}^{qg\rightarrow q + X}(\tau_1^0, \tau_2^0)}$, i.e. use the zero-mass expressions to begin with. This is known as the SACOT scheme \cite{Kniehl:2004fy}. The problem of this choice is that it leads to infinite cross sections towards $P_{\rm T} \rightarrow 0$ due to the ${d\hat{\sigma}^{qg\rightarrow q + X}}/d^3p \sim (\tau_{1,2}^0)^{-n}$  behaviour of the partonic cross sections. In the so-called FONLL scheme \cite{Cacciari:1998it} this is avoided by multiplying the partonic cross section by an ad-hoc damping factor $p_{\rm T}^2/(p_{\rm T}^2 + c^2m^2)$ with $c\sim 5$, which serves to tame the unphysical behaviour at small $p_{\rm T}$. Alternatively, the problematic behaviour can be avoided simply by retaining the kinematics of the $Q\overline{Q}$-pair production which, deep down, is the underlying process we describe. With this physical picture in mind, we define ${d\hat{\sigma}^{Qg\rightarrow Q + X}(\tau_1, \tau_2)} \equiv {d\hat{\sigma}^{qg\rightarrow q + X}(\tau_1, \tau_2)}$ taking $\tau_{1,2} = {m_{\rm T}e^{\mp y}}/({\sqrt{s} x_{2,1}})$ as in FFNS. This leads to well-behaved cross sections in the $P_{\rm T} \rightarrow 0$ limit. We call this the SACOT-$m_{\rm T}$ scheme, as it shares the same underlying idea as the SACOT-$\chi$ scheme in deeply-inelastic scattering \cite{Guzzi:2011ew}.

\begin{figure}[htb!]
\centering
\includegraphics[width=0.88\linewidth]{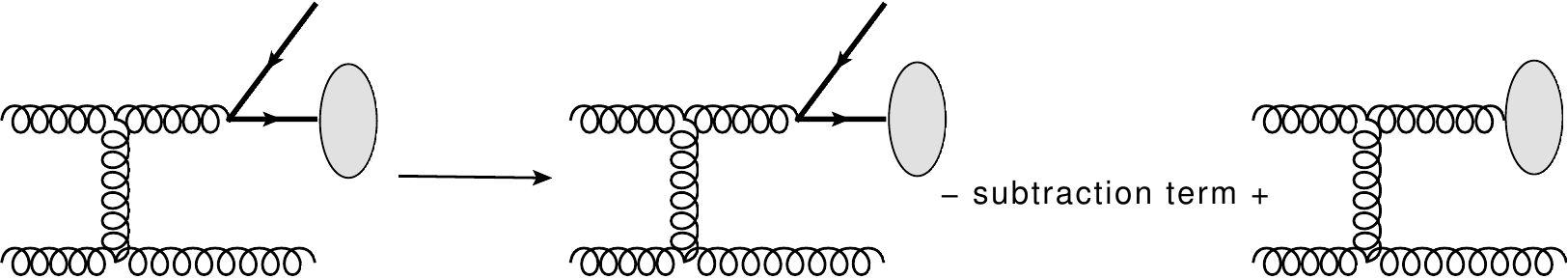}
\caption{Origin of the gluon-fragmentation subprocess illustrated.}
\label{fig:fin}
\end{figure}

Part of the $\log(p_{\rm T}^2/m^2)$ terms in FFNS come also from final-state splittings. As an example, Figure~\ref{fig:fin} shows a diagram in which an outgoing gluon splits into a $Q\overline{Q}$ pair. 
The resulting $\log(p_{\rm T}^2/m^2)$ terms are absorbed into the fragmentation-scale $(\mu_{\mathrm{frag}})$ dependent gluon FFs, $D_{g \rightarrow \mathrm{D}^0}(z,\mu^2_{\rm frag})$, giving rise to gluon-fragmentation contributions,
\vspace{-0.2cm}
\begin{align}
\int \frac{dz}{z}dx_1 dx_2 \, 
            { f_g^{p}(x_1,\mu^2_{\rm fact}) }
            {\frac{d\hat{\sigma}^{gg\rightarrow  g + X}(\tau_1, \tau_2)}{dp_{\rm T} dy}}
            {f_g^{p}(x_2,\mu^2_{\rm fact})}             
            {D_{g \rightarrow \mathrm{D}^0}(z,\mu^2_{\rm frag})}             
        \nonumber \,.
\end{align}
The subtraction term avoiding the double counting is again the same expression, but replacing the gluon FF by its perturbative form, 
\vspace{-0.3cm}
\begin{equation}
D_{g \rightarrow \mathrm{D}^0}(x,\mu^2_{\rm frag}) = \left(\frac{\alpha_s}{2\pi}\right) \log\left(\frac{\mu^2_{\rm frag}}{m^2}\right) \int_x^1 \frac{\mathrm{d}\ell}{\ell} P_{qg}\left(\frac{x}{\ell}\right) D_{Q \rightarrow \mathrm{D}^0}(\ell) + \mathcal{O}(\alpha_s^2)  \nonumber \,. 
\end{equation}
In line with our scheme choice, the $\overline{\rm MS}$ zero-mass matrix elements for $d\hat{\sigma}^{gg\rightarrow  g + X}(\tau_1, \tau_2)$ with the massive expressions for $\tau_{1,2}$, are adopted. Even if the heavy quarks do not explicitly appear in the $gg\rightarrow g+X$ process, the underlying process is also here the $Q\overline{Q}$-pair production. 

With this schematic justification, our ``master formula'' within GM-VFNS is,
\begin{align}
\frac{d\sigma^{pp \rightarrow \mathrm{D}^0 + X}}{dP_{\rm T}dY} = \sum _{ijk}
\int \frac{dz}{z} dx_1 dx_2 
            & { f_i^{p}(x_1,\mu^2_{\rm fact}) }
            { \frac{d\hat{\sigma}^{ij\rightarrow k}(\tau_1, \tau_2, m, \mu^2_{\rm ren}, \mu^2_{\rm fact}, \mu^2_{\rm frag})}{dp_{\rm T} dy} } \\
            & { f_j^{p}(x_2,\mu^2_{\rm fact}) }            
            { D_{k \rightarrow \mathrm{D}^0}(z,\mu^2_{\rm frag}) }             
        \nonumber \,.
\end{align}
Unlike in FFNS, all partonic subprocesses are included and the FFs are scale dependent. In the limit $p_{\rm T} \rightarrow 0$, the partonic cross sections reduce to FFNS, but towards $p_{\rm T} \rightarrow \infty$ they tend to the zero-mass $\overline{\rm MS}$ expressions. Our numerical realization of SACOT-$m_{\rm T}$ scheme is crafted around the Mangano-Nason-Ridolfi code \cite{Mangano:1991jk} for heavy quarks, and the \textsc{incnlo} code \cite{Aversa:1988vb} for zero-mass partons. All terms up to $\mathcal{O}(\alpha_s^3)$ are included. 

\vspace{-0.2cm}
\begin{figure}[htb!]
\centering
\includegraphics[width=0.52\linewidth]{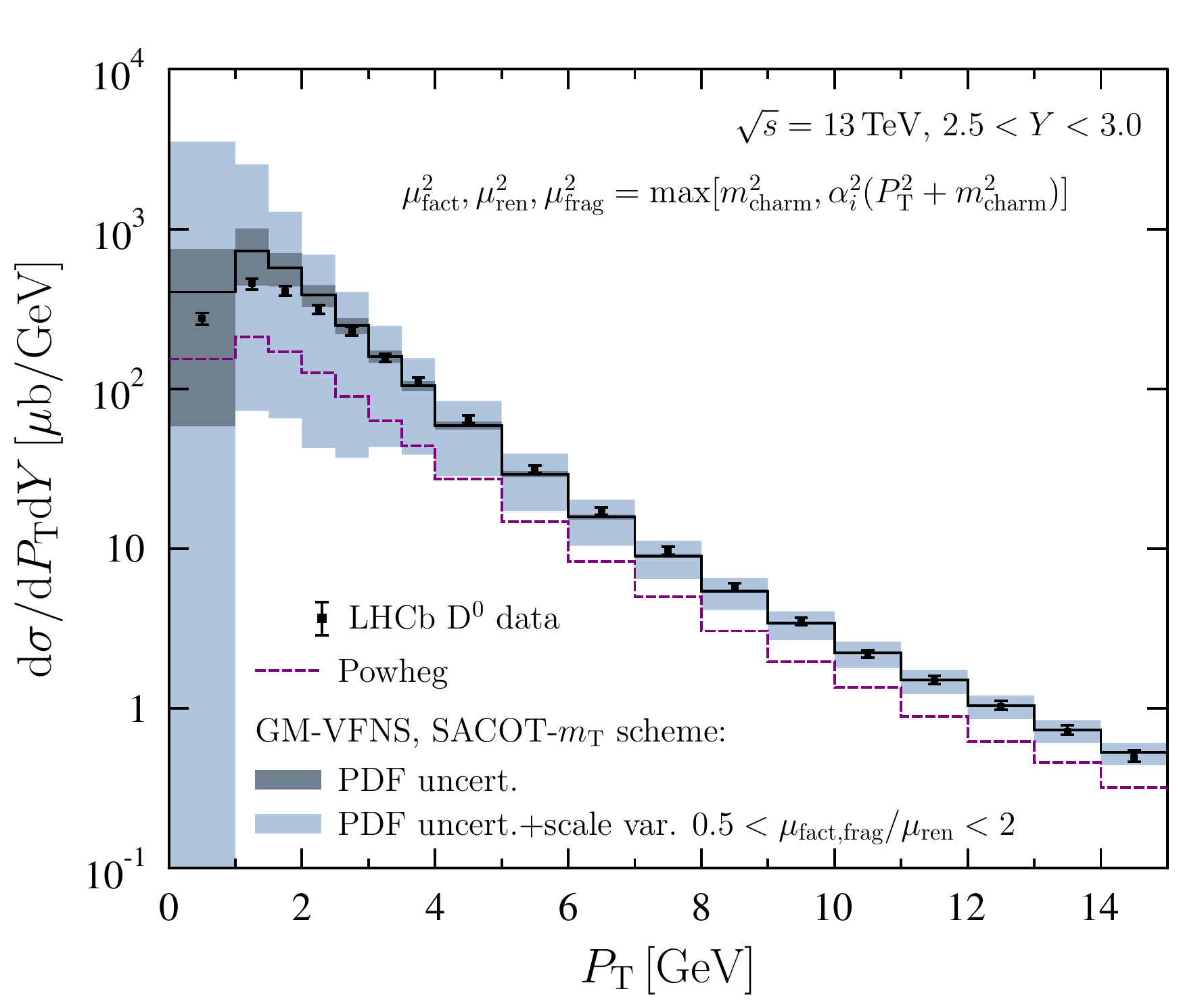}
\includegraphics[width=0.46\linewidth]{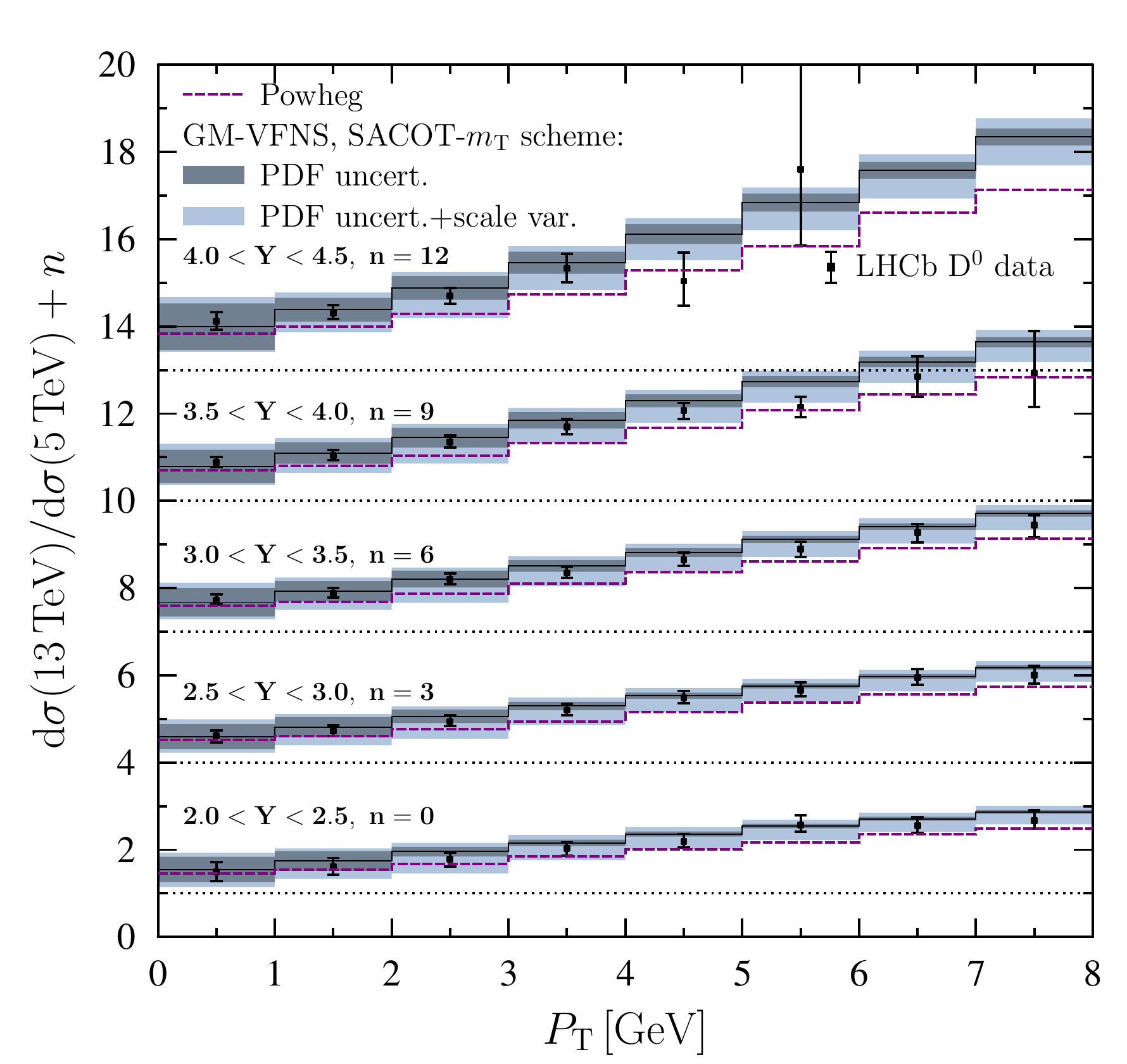}
\vspace{-0.2cm}
\caption{Left: LHCb D$^0$ data \cite{Aaij:2015bpa} in p-p collisions at $\sqrt{s}=13\,{\rm TeV}$ compared to our GM-VFNS calculation and the \textsc{Powheg}+\textsc{Pythia} framework. Right: Ratio between $\sqrt{s}=13\,{\rm TeV}$ and $\sqrt{s}=5\,{\rm TeV}$ data.}
\label{fig:LHCb}
\end{figure} 

\vspace{-0.2cm}
\section{Description of the LHCb D$^0$ data}

\vspace{-0.2cm}
In Figure \ref{fig:LHCb}, we compare the LHCb 13\,TeV D$^0$ data \cite{Aaij:2015bpa} with our GM-VFNS theory calculation. The darker bands show the NNPDF3.1 (pch) \cite{Ball:2017nwa} PDF uncertainty, and the lighter bands combine the scale-variation and PDF errors. We have used here the KKKS08 FFs \cite{Kneesch:2007ey}. The calculation agrees very well with the data though the scale variation leads to a significant uncertainty at small $P_{\rm T}$. We have found that, the contribution from the three FFNS processes, including the subtraction terms, is less than 10\% for $P_{\rm T} \gtrsim 3 \, {\rm GeV}$ and only around 3\% for $P_{\rm T} \gtrsim 10 \, {\rm GeV}$. This demonstrates that the $\log(p_{\rm T}^2/m^2)$ terms in FFNS become quickly the dominant ones and must be resummed as done in GM-VFNS. A comparison using FFNS + parton-shower approach (with the same PDFs) is also presented. Here, we have used the \textsc{Powheg-Box} event generator \cite{Frixione:2007nw} matched to the \textsc{Pythia}~8 \cite{Sjostrand:2014zea} parton shower. We see that the \textsc{Powheg}+\textsc{Pythia} setup has a tendency to underpredict the experimental spectrum, and the ratio between two $\sqrt{s}$ is clearly flatter than the GM-VFNS prediction. We deduce that the leading reason is that by starting only with $\mathrm{c\overline{c}}$ pairs (generated by \textsc{Powheg}) one neglects the contributions in which the parton shower excites the $\mathrm{c\overline{c}}$ pair only later on. In GM-VFNS these are resummed to the scale-dependent FFs and, indeed, e.g. the gluon-to-D contributions can be around 50\% of the total budget. The gluon fragmentation also significantly changes the $x$ regions in which the PDFs are sampled. Thus, the use of FFNS-based computation when using D-meson data to fit GM-VFNS PDFs would inflict a potential bias.

\vspace{-0.3cm}
\section*{Acknowledgments}
\vspace{-0.2cm}

The support by the Academy of Finland Projects 297058 \& 308301, the Carl Zeiss Foundation, and the state of Baden-W\"urttemberg through bwHPC, are acknowledged. We have used the computing facilities of the Finnish IT Center for Science (CSC) in our work.


\vspace{-0.3cm}
\begin{thebibliography}{99}
\vspace{-0.2cm}

\bibitem{Zenaiev:2015rfa}
  O.~Zenaiev {\it et al.},
  Eur.\ Phys.\ J.\ C {\bf 75} (2015)  396.

\vspace{-0.2cm}
  
\bibitem{Gauld:2016kpd}
  R.~Gauld and J.~Rojo,
  Phys.\ Rev.\ Lett.\  {\bf 118} (2017) 072001.

 \vspace{-0.2cm}   
  
\bibitem{Mangano:1991jk}
  M.~L.~Mangano, {\it et al.},
  Nucl.\ Phys.\ B {\bf 373} (1992) 295.
  
\vspace{-0.2cm}  
  
\bibitem{Cacciari:1998it}
  M.~Cacciari, {\it et al.},
  JHEP {\bf 9805} (1998) 007.
 
 \vspace{-0.2cm}  
  
\bibitem{Frixione:2007nw}
  S.~Frixione, {\it et al.},
  JHEP {\bf 0709} (2007) 126.
 
\vspace{-0.2cm}   
 
\bibitem{Helenius:2018uul}
  I.~Helenius and H.~Paukkunen,
  JHEP {\bf 1805} (2018) 196.

\vspace{-0.2cm}    
  
\bibitem{Zenaiev:2016kfl}
  O.~Zenaiev,
  Eur.\ Phys.\ J.\ C {\bf 77} (2017) no.3,  151.
  
\vspace{-0.2cm}   
 
\bibitem{Thorne:2008xf}
  R.~S.~Thorne and W.~K.~Tung,
  arXiv:0809.0714 [hep-ph].
  
\vspace{-0.2cm}  

\bibitem{Kniehl:2004fy}
  B.~A.~Kniehl {\it et al.},
  Phys.\ Rev.\ D {\bf 71} (2005) 014018.
  
\vspace{-0.2cm}  
  
\bibitem{Guzzi:2011ew}
  M.~Guzzi, {\it et al.},
  Phys.\ Rev.\ D {\bf 86} (2012) 053005.
  
\vspace{-0.2cm}  
  
\bibitem{Aversa:1988vb}
  F.~Aversa, {\it et al.},
  Nucl.\ Phys.\ B {\bf 327} (1989) 105.
  
\vspace{-0.2cm}  
  
\bibitem{Aaij:2015bpa}
  R.~Aaij {\it et al.},
  JHEP {\bf 1603} (2016) 159.
  
\vspace{-0.2cm}  
  
\bibitem{Ball:2017nwa}
  R.~D.~Ball {\it et al.},
  Eur.\ Phys.\ J.\ C {\bf 77} (2017) 663.
  
\vspace{-0.2cm}  
  
\bibitem{Kneesch:2007ey}
  T.~Kneesch, {\it et al.},
  Nucl.\ Phys.\ B {\bf 799} (2008) 34.
 
\vspace{-0.2cm} 
 
\bibitem{Sjostrand:2014zea}
  T.~Sjöstrand {\it et al.},
  Comput.\ Phys.\ Commun.\  {\bf 191} (2015) 159.
 
\end{thebibliography}
\end{document}